# Optimizing a Broad Energy High Purity Germanium (BEGe) Detector Operated at Shallow Depth in Abu Dhabi

O. Fawwaz[1], H. Shams[1, +], F. Arneodo[1], A. Di Giovanni[1,2]

[1]*New York University Abu Dhabi, Abu Dhabi, United Arab Emirates*

[2]*Gran Sasso Science Institute, L'Aquila, Italy*

## Abstract

In this work we present the characterization of a Broad Energy Germanium (BEGe) type High Purity Germanium (HPGe) detector, with a carbon fiber entrance window thickness of $0.6\ mm$ and an active area of $6305\ mm^2$, operated at shallow depth (~ 8m) in Abu Dhabi, UAE. A $1.6\ keV$ Full Width Half Maximum (FWHM) was obtained for the $662\ keV$ peak of $^{137}Cs$. A muon veto was applied, reducing the background by 8 % (for energies greater than $100\ keV$). Flushing the volume around the detector endcap with nitrogen gas, to remove radon and thus its progeny, further reduced the background by ~3 %. A thorough analysis for the shaping filter parameters showed that the detector has better resolution at low rise-time values (2 - 5 $\mu s$) especially for low energy gamma (<600keV), keeping the flattop value fixed at 1.1 $\mu s$.

## Article Highlights

- Use of a muon veto and N₂ flushing to improve the detectors' background
- Optimizing shaping filter parameters for the best resolution for detected peaks
- Peak detection with different readout electronics and gain values for different count rates

## Keywords

Broad-Energy Germanium detector, High Purity Germanium, resolution, rise-time, background

## Introduction

Germanium (Ge) detectors have been in use for a long time, however they only became the dominating semiconductor detectors in the 1970's after the production of ultra-pure Ge was developed [1]. High-Purity Germanium (HPGe) has remained dominant as a semiconductor detector because it provides a better resolution compared to other detectors [2], and contains a very low concentration of electrically active defects; a low concentration of defects means that there is not much variation in the local electric-field potential, which makes the purification and crystal growth process easier [3, 4]. HPGe is still the material of choice in the detection of gamma ray photons for its high resolution.

Gamma-ray spectrometry is one of the most commonly used techniques for nuclear radiation characterization and classification. This radio-analytical method analyzes gamma-ray emitting isotopes and is capable of measuring high-level radioactive materials. A variety of detectors are used to serve this purpose, yet HPGe detectors, in their many forms, remain the most common tool for gamma-ray spectroscopy.

In our work here, after installing the detector and shielding, we completed a performance check to verify the detector specifications against manufacturer data. Detector resolution is the most important feature to be tested and optimized. The better the resolution, the better the ability of the detector to separate full energy peaks that are close in energy. When looking at trace concentrations of radionuclides, better resolution also allows for more pronounced peaks. [5]

Our work at "The Astroparticle Laboratory", at New York University Abu Dhabi (NYUAD), described in this paper focuses on improving the sensitivity and resolution of our Broad Energy-HPGe (BEGe) [2, 6–9] detector by applying an active muon veto, flushing the inside of the shield with nitrogen ($N_2$), and thoughtfully adjusting the shaping filter parameters. Changing the rise-time changes the resolution of the peaks, which in-turn allows for achieving the system's highest resolution.

**Experimental Setup**

We have performed a set of measurements using a High-Purity Germanium Broad Energy detector [Canberra BE6530] whose active area is 6305 $mm^2$, operated two levels underground (~8m) in the basement of NYUAD. The detector is equipped with a carbon composite window that is 0.6 mm thick, and a 114.3 mm diameter aluminum end-cap. As a means of cooling, the system is equipped with a CANBERRA Cryo-Cycle II, a "hybrid" cryostat with a $LN_2$ reservoir, and a self-contained cryocooler that condenses the boil-off vapor to maintain the $LN_2$ supply.

Surrounding the detector is a composite passive shield consisting of several layers (from outside to inside): 9.5 mm thick low-carbon steel, 15 cm of ordinary low-background lead, 2.5 cm of low activity $^{210}$Pb (activity of about 20 Bq/kg; used to stop the bremsstrahlung from the outer lead volume), 1 mm of low-background tin (used in the molding of the lead layer, as well as an aid in adhering the lead to the copper, and as an absorber), and 1.5 mm of high-purity copper (graded liner to stop lead K-shell x-rays). Two large area plastic scintillators (BC408 Saint-Gobain) are added as an active muon veto; both are placed horizontally, one above and the other is below the detector outside the lead shielding (see Fig. 1). Additionally, nitrogen flushing through a gas port leading into the endcap is applied to the detector as a means to diminish radon (both $^{220}Rn$ and $^{222}Rn$) and its progeny (e.g. $^{214}$Pb, $^{214}$Bi, $^{212}$Pb) [10].

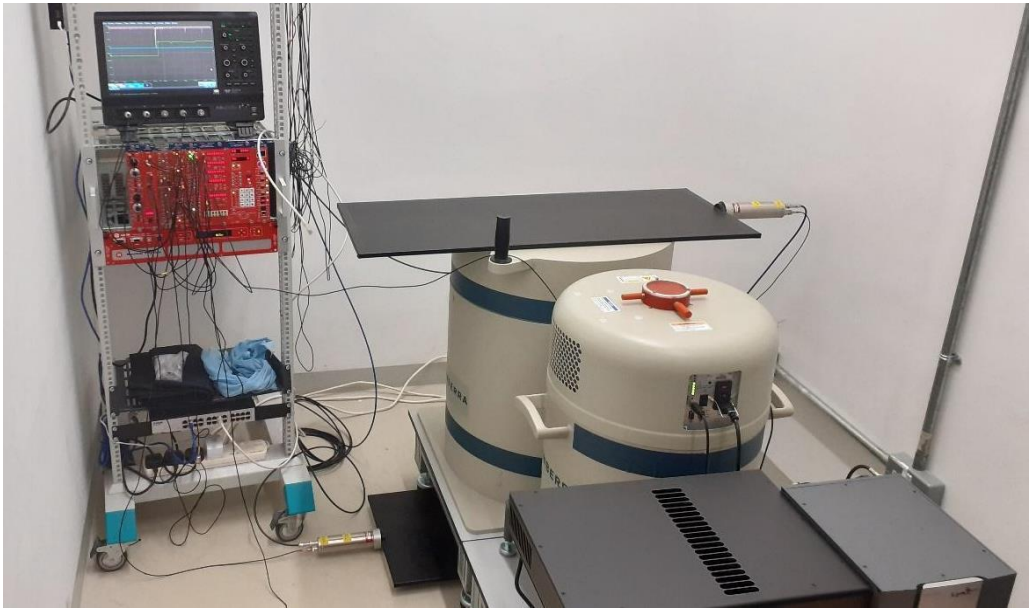

Fig. 1 The experimental set-up at New York University Abu Dhabi

Data acquisition and signal processing was done using the digitizer provided with the detector from Mirion Technologies, namely, a LYNX MCA digitizer. A CAEN NIM crate and modules are used to acquire and process the signal from the veto scintillators, and the output is used to set an anti-coincidence at the LYNX, which is in-turn controlled and visualized by Genie 2000, a software environment used for gamma and alpha spectrometry data analysis. The modules used in the veto system are the following (in order): Quad Linear Fan-IN Fan-OUT [Mod. N625], Low Threshold Discriminator [8 CH LTD - Mod. N844], 3-Fold Logic Unit [Mod. N405] (to create a coincidence between the two scintillators), Quad Scalar and Preset counter Timer [Mod. N1145], Dual Timer [Mod. N93B] (to produce a 30 μs square signal which is used to stop the acquisition at the LYNX) and NIM-TTL-NIM adapter [Mod. 89] to fit the LYNX input specifications for a signal.

**Calibration**

The detector was calibrated using a NORM (Naturally Occurring Radioactive Material) sample [11], a scale formed in a pipe at an oil field, containing $^{226}$Ra and its progeny; a sample spectrum is presented in Fig. 2; for the sake of a comparison the same spectrum as obtained by a ORTEC, 3×3 inch – 905 Series, NaI(Tl) scintillation radiation detector coupled to a digiBASE 14-Pin PMT Base with Integrated Bias Supply is added in the same figure. A nuclide library was created and used for the calibration. A 6.72 × gain was used to make sure that the 2.614 MeV gamma-ray from $^{208}$Tl is visible in the spectrum.

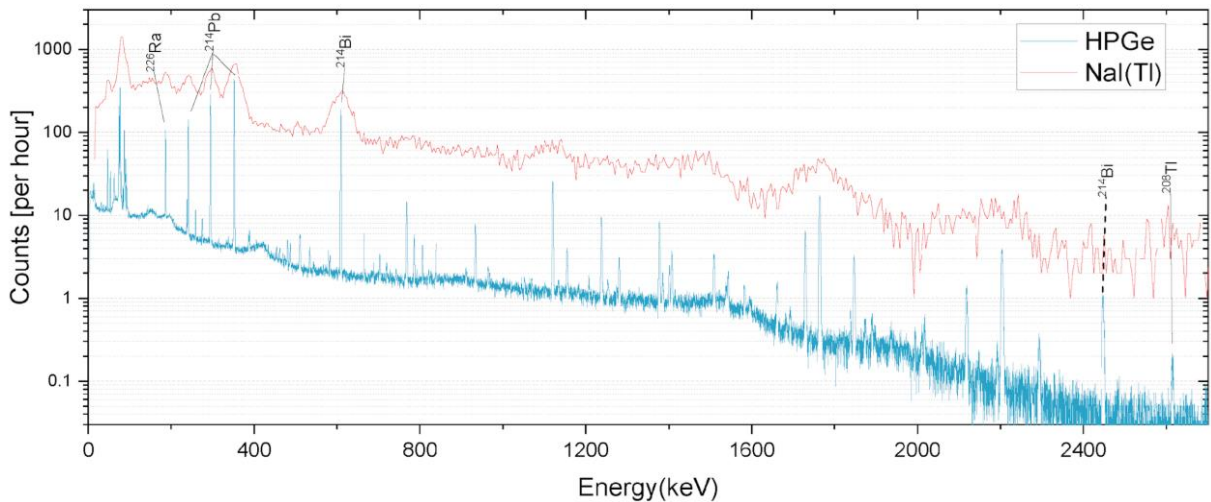

Fig. 2 Spectrum of a NORM scale sample, showing $^{226}$Ra, taken with the Ge detector(blue), and a spectrum of the same sample taken with a 3x3 inch NaI(Tl) scintillating crystal (red). (Acquisition time is 2 days)

**Muon Veto and N$_2$ flushing**

The detector is located two levels underground. The local flux of cosmic rays at that location has been measured with the same technique used in [12], finding an intensity of about 40 % with respect to the surface. In order to reduce the cosmic-ray induced background count rate we implemented a veto system with two Scionix EJ200 plastic scintillators, each 1000 mm long with a cross-section of 400x20 mm, wrapped in special reflective material and light tight vinyl, coupled to a 38 mm diameter Hamamatsu R980 photomultiplier tube (PMT). When a coincidence happens, a square signal with a pulse width of 30 μs is generated by the Dual Timer. That signal is then fed to the LYNX veto input. This system lowers the background count rate for energies larger than 100 keV by 8% and the 511 keV annihilation peak by 14 % (Fig. 3).

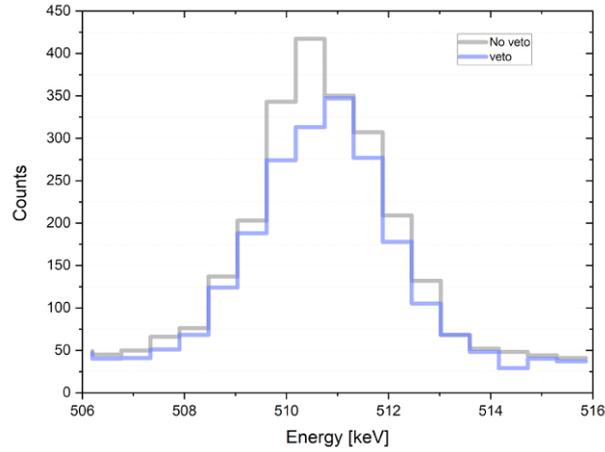

Fig. 3  511 keV full energy peak with and without using the muon veto (2 days acquisition time)

To further reduce the background count rate, we flushed the interior of the shield with $N_2$ to diminish radon progeny build-up. Fig. 4 shows that the nitrogen flushing diminishes the $^{214}$Pb 351.9 keV peak almost entirely. The total number of background counts above 100 keV is reduced by 3 %.

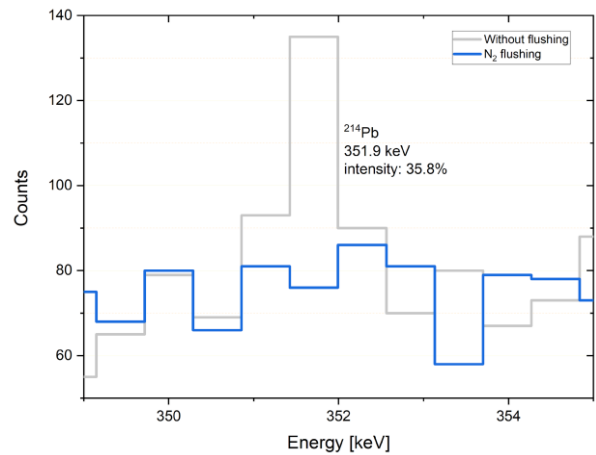

Fig. 4 $^{214}$Pb full energy peak without (grey) and with $N_2$ flushing (blue) (2 days acquisition time)

### Rise-time and flattop

The effect of the rise-time setting of the shaping on the resolution is described in literature [13–15], and different values for the rise-time were reported by various publications [13, 15–18], while the one recommended by Mirion, for our detector, is 10 µs. We have conducted a detailed study of this effect on the resolution for a wide range of energies and rise-time values. Point-like calibration sources were used ($^{57}$Co, $^{60}$Co, $^{137}$Cs, $^{155}$Eu, and $^{22}$Na) for which the live time of acquisition was 1500 s, to ensure the minimum area (number of counts) in any selected full energy peak is more than ten thousand counts. The acquisition was automated, using ReXX, the environment provided by Mirion, which was useful for controlling the acquisition and the analysis from the command line. For a fixed flattop, the rise-time values were varied from 1 μs to 12 μs, and for each setting the measurement was repeated 10 times. For energies less than 1 MeV, low rise-time values (3 μs – 4 μs) were favorable, but the most significant effect on the FWHM was for low energy peaks, below 500 keV, as shown in Fig. 5. The rise-time values for high energies were extended to 20 μs (see Fig. 6).

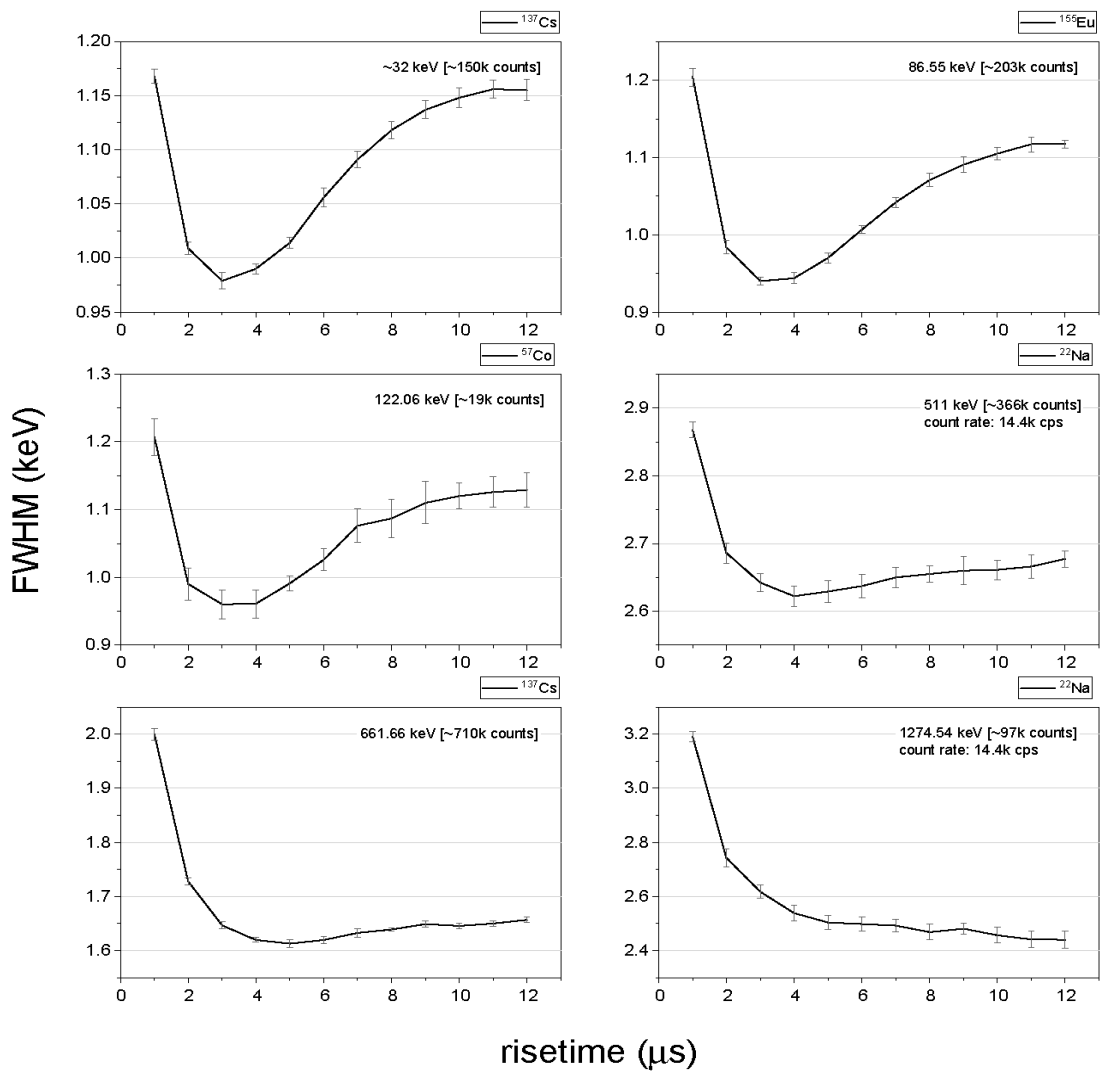

Fig. 5 FWHM versus rise-time (at a fixed flattop of 1.1 μs) for various peaks using $^{57}$Co, $^{155}$Eu, $^{22}$Na, and $^{137}$Cs point-like calibration sources, best rise-time values were: 3 μs for <100 keV, 4 μs between 100 and 500 keV, 5 μs for the 662 keV full energy peak. Average area (number of counts) in each case is reported, and the error bars represent the standard deviation in each case

To investigate the effect of count rate, lead sheets were used along with a $^{22}$Na calibration source to reduce the count rate to less than one third of the count rate without the sheets. The results show a slight variation for the 511 keV peak, indicating a better resolution at 5 μs rather than 4 μs, however, given the large uncertainty at the low count rate it is not so clear; on the other hand the 1274 keV full energy peak showed a very similar behavior to that of the high count rate, and results are shown in Fig. 7. This investigation was then repeated for a much lower count rate, 8.5 counts per second (cps) using the low activity NORM sample, however a lower range of values and number of repetitions were used. Although we tried to get at least 10k counts for each selected peak, which took a large amount of time (averaging ~6 hours for each data point/measurement), the obtained areas were much smaller than in the point source case, which explains the large uncertainties seen in Fig. 8.

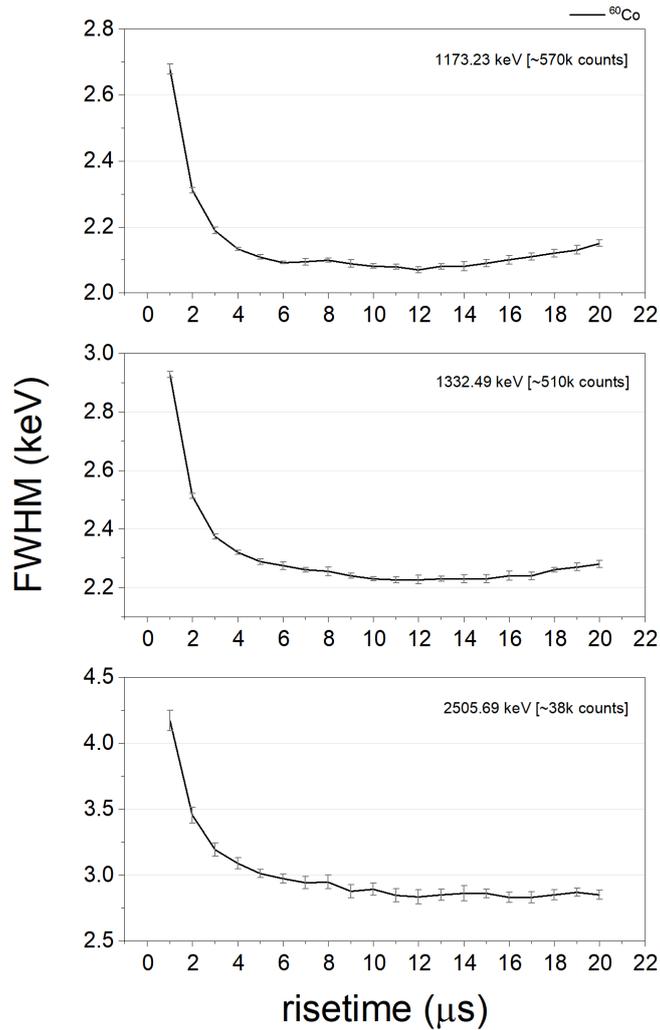

Fig. 6 FWHM versus rise-time (at a fixed flattop of 1.1 μs) for the three full energy peaks of $^{60}$Co showing the rise-time of 12 μs to be more suitable. The average area (number of counts) in each case is reported, and the error bars represent the standard deviation in each case

These results were obtained using the LYNX Digitizer provided by Mirion. In order to see the influence of the readout electronics, the investigation was also done using a CAEN digitizer (DT5780). In Fig. 9 the results for three full energy peaks of a mixed $^{155}Eu$-$^{22}Na$ calibration source are shown. A better resolution appears at lower rise-time values, but here this is also the case for the high energy.

Additionally, we varied the flattop value from 0.1 μs to 3 μs in steps of 0.1 μs, with 10 runs for each flattop value, at a fixed rise-time of 4 μs. This data (Fig. 10) shows that the FWHM fluctuates for flattop values less than 1.1 μs and remains rather stable all the way up to the maximum tested flattop of 3 μs. Note that for full energy peaks 662 keV and above, there are "missing" data points; this is due to the inability to detect the full energy peaks at these flattop values.

The above work was done at a gain of 6.72. To test gain dependence, a higher gain of 26.9 was analyzed and compared. With a higher gain a smaller energy range would be covered; thus a full energy peak would appear shifted to the right side of the spectrum window (with a higher bin location) if compared to a lower gain setting. Fig. 11 displays the results for the full energy peak of $^{155}$Eu at 86.5 keV , showing a very small variation of the FWHM dependence on the rise-time value, while the FWHM of the 662 keV full energy peak behaves, for increasing rise-times, like high energy lines in the lower gain setting (Fig. 12).

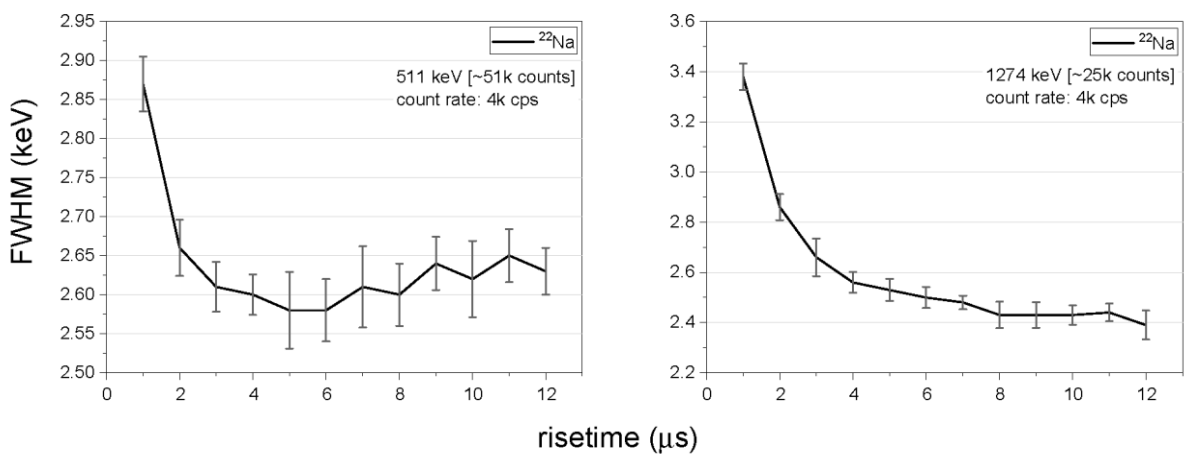

Fig. 7 FWHM versus rise-time for $^{22}$Na full energy peaks at low count rate (4k cps), showing similar behavior for the same full energy peaks at higher count rates presented in Fig. 5

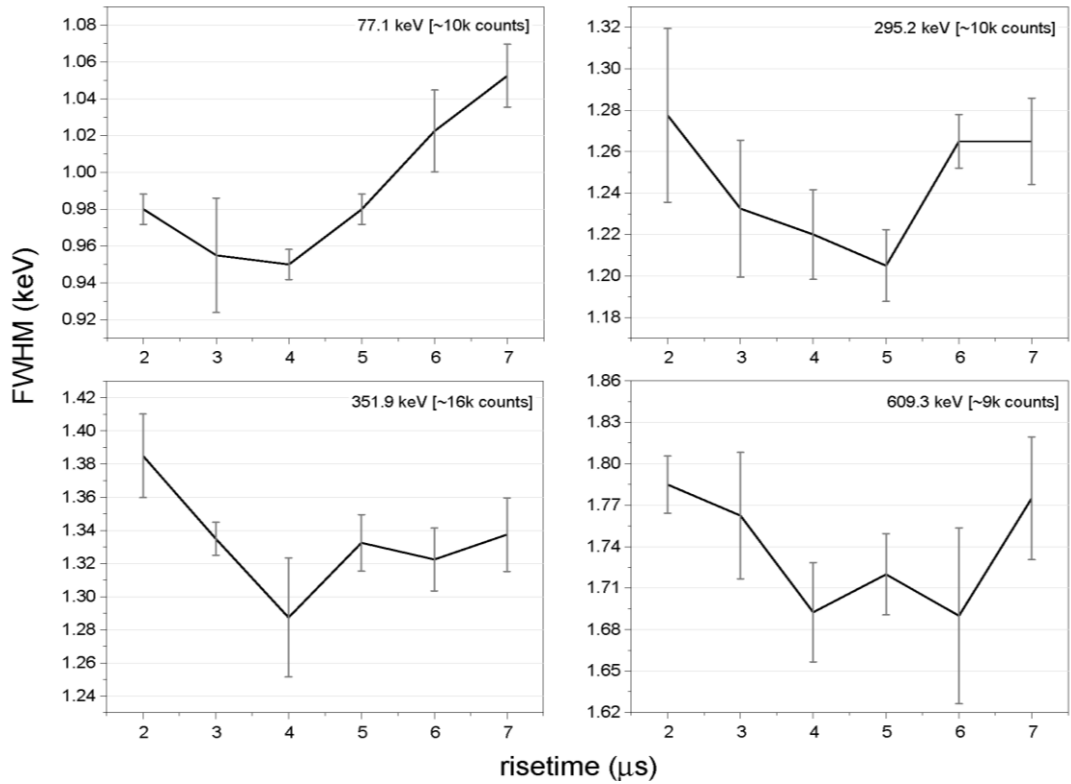

Fig. 8 FWHM versus rise-time (at a fixed flattop of 1.1 μs) for four full energy peaks that appear in the $^{226}$Ra spectrum. The average area (number of counts) in each case is reported, and the error bars represent the standard deviation in each case

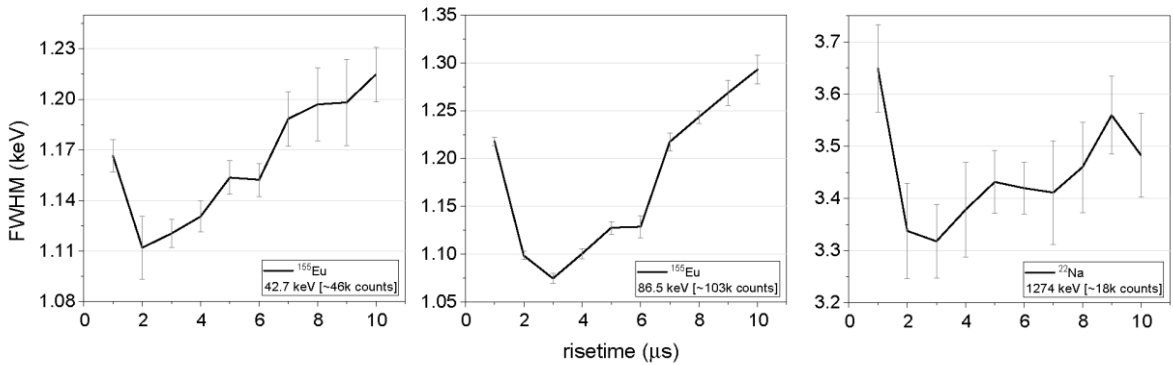

Fig. 9 FWHM versus rise-time plots using a CAEN digitizer, a mixed $^{155}$Eu-$^{22}$Na calibration source was used

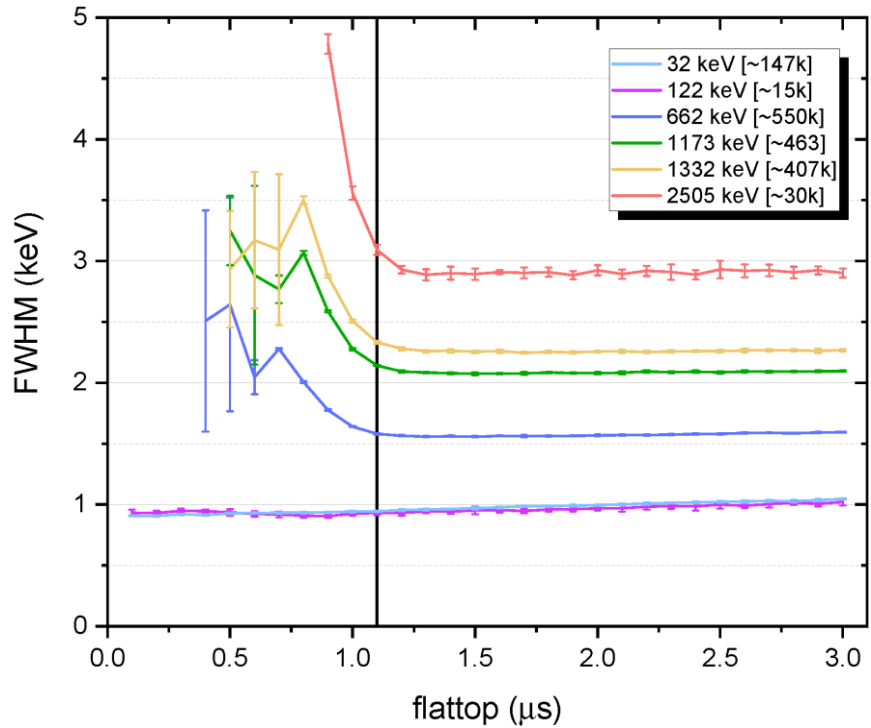

Fig. 10 FWHM versus flattop (at a fixed rise-time of 4 µs) for different energies. The average area (number of counts) in each case is reported in square brackets next to the energy value, and the error bars represent the standard deviation in each case

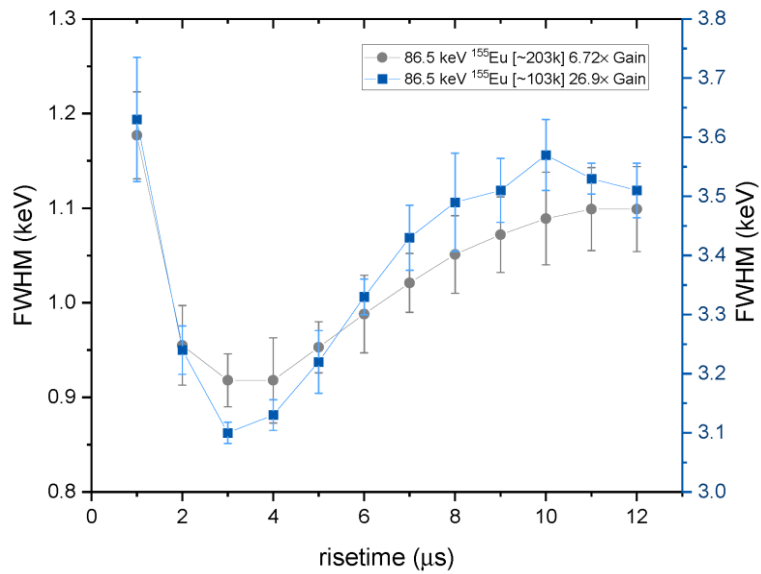

Fig. 11 Increasing the gain creates slight change at low energies, e.g. for the 86.5 keV full energy peak the higher gain (in blue) shows a better resolution at 3 µs

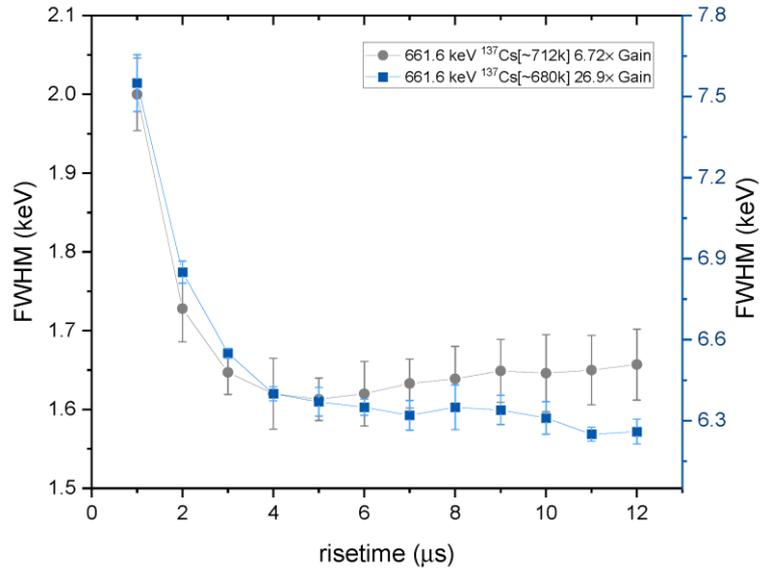

Fig. 12 Increasing the gain shows that the FWHM at the 662 keV full energy peak no longer shows the minimum at 5 μs rise-time, rather favoring larger values, similar to the full energy peaks with energies >1000 keV in the lower gain setting

## Conclusion

In this study we present the effect of the shaping filter parameters on the resolution of a BEGe detector. The investigation was done for different energies ranging from 32 keV up to 2.5 MeV, and we used two different digitizers, a LYNX MCA digitizer from Mirion Technologies and another from CAEN (model DT5780). A muon veto was implemented along with $N_2$ flushing within the endcap surrounding the Ge crystal to reduce the background count rates by ~11%.

Several count rates were investigated, using varying activity point sources, by using lead sheets to reduce the count rate of a $^{22}$Na calibration source and a low activity NORM sample. High and low gain values (6.72 and 26.9) were used to check for dependence; this resulted in a higher gain, for the same energy, requiring a slightly higher risetime value to achieve a better resolution. The selection of the rise-time value depends on the energy range of interest; for energies less than 600 keV, low rise-time values (2-5 μs) result in a better resolution, whereas for higher energies higher rise-times are necessary.

By implementing the muon veto, flushing the detector volume with $N_2$, and bettering the relevant rise-time, we were able to optimize the BEGe detector and achieve a better resolution.

## Acknowledgements

We would like to thank the Core Technology Platforms at New York University Abu Dhabi, represented by its Executive Director, Mr. Reza Rowshan and his team, for their precious support and funding of this study.

We would also like to extend our thanks to Dr. Philip Rodenbough, Adjunct Lecturer (Scientific Writing) at New York University Abu Dhabi, as well as Isaac Sarnoff, Physics PhD student at New York University Abu Dhabi, for fine-tuning the language and structure of the manuscript.